# Theory of Exceptional Points of Degeneracy in Uniform Coupled-Waveguides and Balance of Gain and Loss

Mohamed A. K. Othman and Filippo Capolino

*Abstract*—We present a transmission line theory of exceptional points of degeneracy (EPD) in coupled-mode guiding structures, i.e., a theory that illustrates the characteristics of coupled electromagnetic modes under a special dispersion degeneracy condition, yet unexplored in the contest of gain and loss. We demonstrate the concept of Parity-Time ($\mathcal{PT}$)-symmetry in coupled uniform waveguides with balanced and symmetric gain and loss and how this condition is associated with a second order EPD. We show that by introducing gain into naturally lossy structures provides for the conditions whereby exceptional points of non-Hermitian degeneracies can be manifested, such as in $\mathcal{PT}$-symmetric structures. Furthermore, we also demonstrate that $\mathcal{PT}$-symmetry, despite being the method often suggested for obtaining non-Hermitian degeneracies at optical frequencies, is not a necessary condition and indeed we show that EPD can be obtained with broken topological symmetry in uniform TLs. Operating near such special degeneracy conditions leads to potential performance enhancement in a variety of microwave and optical resonators, and devices such as distributed oscillators, including lasers, amplifiers, radiating arrays, pulse compressors, and *Q*-switching sensors.

## I. INTRODUCTION

EVOLUTION of a general dynamical system described by an eigenvalue problem that depends on a certain parameter can encounter points at which two, or more, physical eigenstates coalesce into one; those are called exceptional points of degeneracy (EPD). Despite in certain physics literature such condition is simply referred to as EP that may be ambiguous to some other research communities, here we include a "D" as "degeneracy" in the acronym to specify the kind of points we are referring to. At those EPDs, the system would allow degenerate states and the corresponding state matrix cannot be diagonalized. It follows that independent basis for representing the electromagnetic states in a waveguide can only be found using generalized eigenvectors leading to non-periodic solutions in space as we will elaborate in the following. An EPD represents a point of singularity in the spectrum of eigenmodes [1], [2] that has also a degeneracy of *system eigenvectors*. Degeneracy conditions are not common but can be found or engineered in many wave guiding structures because they can be very useful to conceive a variety of devices. The simplest degeneracy can be found at the transmission band edge of any periodic guiding structure, where a regular band edge (RBE) dictates that forward and backward modes coalesce at the band edge (i.e., the separation between pass and stop bands). There, group velocity vanishes in a lossless system. Other trivial degeneracies occur at the cutoff frequency of waveguides or at zero frequency. Here instead we discuss more elaborated and useful degeneracy conditions. Periodic structures in particular can offer interesting degeneracies related to the electromagnetic band gap existing in the spectrum of modes, which would not be attainable in uniform waveguides. A *fourth order EPD*, for instance, occurs when all four independent Bloch eigenvectors in lossless periodically coupled waveguides coalesce and form one single eigenvector [3]–[5] at the band edge. This degenerate band edge (DBE) condition is the basis for a huge enhancement of local density of states, giant resonances in crystals [5]–[7], in coupled transmission lines [8]–[10], and in ladder circuits [11] and consequently a possible enhancement of gain in active devices comprising DBE structures [7], [10]. However, the focus of this paper is on second order EPDs in *uniform* waveguides with *balanced gain and loss*.

We point out EPDs cannot occur in systems whose evolution is described by a Hermitian matrix (see Section III). Thus, degeneracy conditions occur in systems that require non-Hermiticity [4], [7], [12], [13] such as the periodic systems described above. In this paper, we investigate unconventional *non-Hermitian* EPDs [2], [14], [15]; occurring in uniform guiding structures whose wave dynamics are represented by two coupled transmission lines (CTLs) in the presence of both *gain and loss*. The first provided example of such peculiar EPD manifests in guiding structures supporting the so-called Parity–Time ($\mathcal{PT}$)-symmetry condition [2], [16]–[19], implying that even though the system matrix that describes field evolution along the CTL is not Hermitian, the system eigenvalues can still be real [2] when perfect gain and loss symmetric balance is in place. (We recall that a sufficient condition for a matrix to have

This work has been supported by AFOSR Grant FA9550-15-1-0280 and also by AFOSR MURI Grant FA9550-12-1-0489 administered through the University of New Mexico.

M. A. K. Othman and F. Capolino are with the University of California, Irvine, CA 92697 USA. (e-mail: mothman@uci.edu, f.capolino@uci.edu).

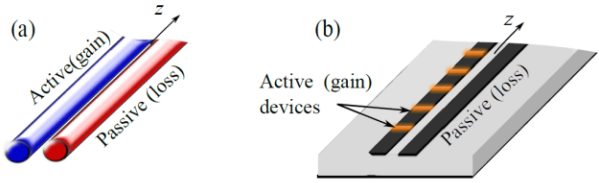

Fig. 1. Example geometries of guided wave structures that potentially support a non-Hermitian EPD in the dispersion diagram. (a) and (b) represents coupled waveguide structures at optical and RF frequencies, respectively, with balanced gain and loss. Waveguides can reach gain and loss balance based on perfect symmetry (the so called $\mathcal{PT}$-symmetry condition) as in (a) that leads to a second order EPD. The geometry in (b) composed with two lossy and coupled microstrips for example, can satisfy the gain and loss balance condition and hence an EPD without the necessity of symmetric gain and loss (i.e., without $\mathcal{PT}$-symmetry).

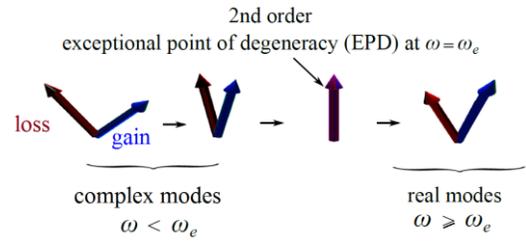

Fig. 2. Schematic representation of the eigenvectors of the coupled waveguides near an EPD as the frequency approaches $\omega_e$, the EPD radian frequency. Evolution of the two eigenvectors near a second order EPD (the other two modes, of a total of four modes, exhibit the same behavior) when the system satisfies $\mathcal{PT}$-symmetry (i.e., perfect gain and loss symmetry). When EPD is achieved without gain and loss symmetry (i.e., without $\mathcal{PT}$-symmetry), modes do not have necessary purely real wavenumbers for $\omega > \omega_e$.

real eigenvalues is to be Hermitian.) Remarkably, when the "sufficient" condition is not satisfied, interestingly a wide class of non-Hermitian system matrices can still possess entirely real spectra. Among these are systems obeying $\mathcal{PT}$-symmetry [1], [2], [16]–[20].

Real spectra of non-Hermitian operators (such as those obeying $\mathcal{PT}$-symmetry) and the occurrence of EPDs have opened new horizons on several fronts of physics, including quantum field theories and quantum interactions [2], [15]. The notion of *phase transition* [1] is often used to indicate the exceptional point at which spontaneous $\mathcal{PT}$-symmetry breaking occurs, that is the point beyond which (depending on which variable is changing, e.g., frequency, gain and loss parameters, coupling, etc.) the spectrum ceases to be real [16], [21], [22]. The concepts of $\mathcal{PT}$-symmetry have been employed in optics; and interesting properties have been observed in coupled waveguides with $\mathcal{PT}$-symmetry when the system's refractive index obeys $n(x) = n^*(-x)$ where $x$ is a coordinate in the system [16], [17], [23], [24]. In the radio frequency region, there has been some studies on EPDs and $\mathcal{PT}$-symmetry in lumped circuits [25]–[27] and in a microwave cavity [18]. However, a more comprehensive investigation must be carried out inspired also by applications and intriguing physics under development in optics.

Furthermore, we show that such $\mathcal{PT}$-symmetry is not necessary at all to achieve an EPD. Indeed, we show an important exemplary case where both TLs have losses, and gain is introduced in only one TL, hence breaking the topological symmetry. Such a CTL system can still exhibit EPD, paving the way to possible future electronic devices based on this peculiar physical effect. Properties based on the use of the EPD include unidirectional propagation [28], [29], coherent perfect absorption [30], [31], low-threshold lasing [32]–[34], as well as pronounced nonlinear soliton propagation [21], [35], [36], sensors [37], and possibly solid state distributed oscillators with the possibility to radiate (a form of gain and loss balance condition defined in the following section).

## II. MOTIVATION AND COUPLED WAVEGUIDES WITH EPDs

In this paper, we develop the theory that reveals the origin of the EPDs in coupled waveguides by adopting a simple CTL model that describes wave propagation in structures such as those in Figs. 1 (a) and (b), as examples of coupled waveguides with perfect gain and loss compensation. The perfect gain and loss symmetry shown in Fig. 1(a) represents what has been referred to as Parity-Time ($\mathcal{PT}$) symmetry, however we will show that this is not a necessary condition to achieve an EPD in practical terms. The evolution of the eigenvectors of in a CTL system near a second order EPD is schematically represented in Fig. 2(a); where two eigenvectors of the system coalesce. In a $\mathcal{PT}$-symmetric system the EPD denotes a system transition from a broken $\mathcal{PT}$-symmetry phase to a perfect $\mathcal{PT}$-symmetry phase regime [17], [23], [28]; when a system parameter is varied. As we will show, in more general terms, without imposing perfect $\mathcal{PT}$-symmetry, when gain and loss balance is achieved so that an EPD occurs, this EPD does not necessarily separate complex modes from real modes, and in this case the dynamics is more involved as shown in Section V.

In this paper, we show for the first time a CTL approach to reveal the features of $\mathcal{PT}$-symmetry in coupled waveguides, and also how these features are related to EPD and to the previously investigated DBE [4], [7], [9]. Furthermore, in Section V we also show how the $\mathcal{PT}$-symmetry is not necessary to achieve EPD opening the path to many other schemes to achieve EPD. Accordingly, we define "gain and loss balance condition" as the condition that guarantees the occurrence of an EPD in such CTL system (without necessarily assuming a perfect gain and loss topological symmetry). Analyses of disorder in the coupled waveguides in the form of imperfect coupling, symmetry breaking or presence of unbalanced gain and losses provide a physical insight into how points of degeneracy are perturbed.

We adopt in this paper an example based on microstrip CTLs whose parameters are given in Appendix A. However, note that the conclusions drawn can be extended to many other geometries or guiding structures since our formalism is general;

operating from RF to optical frequencies for which degeneracies can be found. We limit our discussion in this paper to reciprocal and linear coupled waveguide systems. The rest of the paper is organized as follows. In Section III we develop the theory of CTLs, and of the characteristics of exceptional points in uniform structures therein. In Section IV we investigate the effect of losses and coupling on the dispersion diagram and the properties of the $\mathcal{PT}$-symmetry structures. In Section V we show that EPD is achieved without the need of $\mathcal{PT}$-symmetry. Throughout this paper, we assume time-harmonic fields varying as $e^{j\omega t}$ that is not carried over for simplicity.

III. SYSTEM DESCRIPTION OF CTLS

We consider a waveguide where two modal fields coexist and are able to propagate (or attenuate) along both the positive and negative $z$-directions. The following procedure can also describe more than two modes, up to arbitrary $N$ modes. However, here we only focus on two uniform coupled TLs that pertain to two coupled waveguide structures as shown in Fig. 1; because to achieve the degeneracy of order two described here, only two modes (four if we consider the $\pm k$ symmetry, where $k$ is a wavenumber) are sufficient. Waveguides can be unbounded in the transverse-to-$z$ direction, and it may be made of uniform [Fig. 1(a) and Fig. 3(a)] or quasi-uniform [Fig. 1(b) and Fig. 3(b)] that can be approximated as a uniform system. Let $\mathbf{E}_t(\mathbf{r})$ and $\mathbf{H}_t(\mathbf{r})$ be the transverse components of the electric and magnetic fields relative to two modes supported by the uniform waveguide. For simplicity, we assume that separation of variable in each segment is applicable, and the transverse electric field is represented as

$$\mathbf{E}_{t,1}(\mathbf{r}) = \mathbf{e}_1(\boldsymbol{\rho})V_1(z) + \mathbf{e}_2(\boldsymbol{\rho})V_2(z), \quad (1)$$

where $\mathbf{r} = \boldsymbol{\rho} + z\hat{\mathbf{z}}$, and $\boldsymbol{\rho} = x\hat{\mathbf{x}} + y\hat{\mathbf{y}}$. Analogously we have for the magnetic field

$$\mathbf{H}_{t,1}(\mathbf{r}) = \mathbf{h}_1(\boldsymbol{\rho})I_1(z) + \mathbf{h}_2(\boldsymbol{\rho})I_2(z), \quad (2)$$

where $\mathbf{e}_n(\boldsymbol{\rho})$ and $\mathbf{h}_n(\boldsymbol{\rho})$, with $n = 1,2$, are the electric and magnetic modal eigenfunctions, respectively and $V_n$ and $I_n$ are the amplitudes of those fields that describe the evolution of electromagnetic waves along the $z$-direction (refer to [38], [39] for description of the coupled-mode theory and to [40], [41] for a precise transmission formalism of guided EM waves in waveguides). We can assume for simplicity that $\mathbf{e}_1(\boldsymbol{\rho})$ and $\mathbf{e}_2(\boldsymbol{\rho})$ are orthonormal eigenfunctions. That means they are orthogonal with an inner product defined in the cross section with a unitary norm), and consequently the same for $\mathbf{h}_1(\boldsymbol{\rho})$ and $\mathbf{h}_2(\boldsymbol{\rho})$. This is true for representing uncoupled modes, for example waveguides that are not coupled or far away from each other such that coupling is ignored. When waveguides are coupled (for example at interfaces of anisotropic materials, or in coupled waveguide geometries in Fig. 1), fields can still be represented by those eigenfunctions, such as the case for even and odd field distributions for example. In coupled waveguides we can still define modes that are mutually orthogonal, that however consist of *equivalent* voltage and current terms along the $z$-direction. Accordingly, utilizing those eigenfunctions, the guided EM fields can be well-described by the evolution of their transverse amplitudes (equivalently voltages and currents) resulting in TL equations that will be discussed in the next subsection. As a result, for the structures under consideration as those in Fig. 1, we model wave propagation along the $z$-direction for the two modes of interest as CTLs. Therefore, based on (1), and (2) we consider the amplitudes of the electric and magnetic fields as equivalent voltage and current phasor 2-dimensional vectors are defined as $\mathbf{V}(z) = [V_1(z) \ \ V_2(z)]^T$ and $\mathbf{I}(z) = [I_1(z) \ \ I_2(z)]^T$. Accordingly, the examples in Figs. 1 (a), and (b) can be well represented by the respective cases in Fig. 3(a), and (b) respectively, including coupling coefficients per unit length.

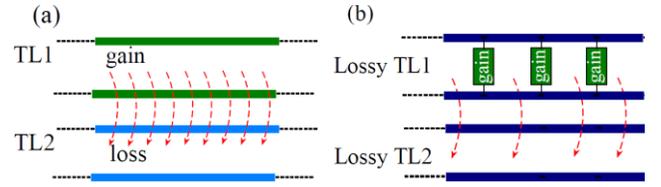

Fig. 3. Configurations of CTLs that may exhibit exceptional points of degeneracy. (a) Uniform CTL with distributed and symmetric gain and loss. (b) Uniform CTL with distributed loss in both TLs and gain in only one that does not obey topological symmetry. In each case, gain and/or loss can be lumped, but densely enough to be approximated as a uniform distribution.

A. *Uniform Coupled-Transmission Lines*

The equations for CTLs consisting of 2 TLs are derived based on the per-unit length distributed parameters and using the matrix notation [42], [43]. We have the following system of coupled first order differential equations representing the fields in uniform transmission lines

$$\frac{\partial \mathbf{V}(z)}{\partial z} = -\underline{\underline{\mathbf{Z}}}\,\mathbf{I}(z), \qquad \frac{\partial \mathbf{I}(z)}{\partial z} = -\underline{\underline{\mathbf{Y}}}\,\mathbf{V}(z) \quad (3)$$

where $\underline{\underline{\mathbf{Z}}}$ and $\underline{\underline{\mathbf{Y}}}$ are the series impedance and shunt admittance matrices describing the per unit parameters of the CTL defined as

$$\underline{\underline{\mathbf{Z}}} = j\omega\underline{\underline{\mathbf{L}}} + \underline{\underline{\mathbf{R}}}, \qquad \underline{\underline{\mathbf{Y}}} = j\omega\underline{\underline{\mathbf{C}}} + \underline{\underline{\mathbf{G}}} \quad (4)$$

where for example, the inductance and capacitance are 2×2 symmetric and positive definite matrices [42], [43], given in the form

$$\underline{\underline{\mathbf{L}}} = \begin{pmatrix} L_{11} & L_{12} \\ L_{21} & L_{22} \end{pmatrix}, \quad \underline{\underline{\mathbf{C}}} = \begin{pmatrix} C_{11} & C_{12} \\ C_{21} & C_{22} \end{pmatrix} \quad (5)$$

The 2×2 symmetric series resistance $\underline{\underline{\mathbf{R}}}$ and shunt conductance $\underline{\underline{\mathbf{G}}}$ matrices account for losses and also for small-signal linear gain, accounted for by negative resistance or conductance. Note that $\underline{\underline{\mathbf{R}}}$ and $\underline{\underline{\mathbf{G}}}$ are positive-definite matrices if and only if they represent only losses [42], [43]. Cutoff conditions could be modeled by resonant series and shunt reactive elements as was done in [10], [12].

To cast the two telegrapher equations in a Cauchy type first order partial differential equation, it is convenient to define the four-dimensional state vector

$$\mathbf{\Psi}(z) = \begin{bmatrix} V_1(z) & V_2(z) & I_1(z) & I_2(z) \end{bmatrix}^T \quad (6)$$

that comprises voltages and currents at a coordinate $z$ in the CTL. Therefore the first order differential equations for the CTL is in written as [4], [10], [12]

$$\frac{\partial}{\partial z}\mathbf{\Psi}(z) = -j\underline{\mathbf{M}}\mathbf{\Psi}(z) \quad (7)$$

where $\underline{\mathbf{M}}$ is a 4×4 CTL system matrix. In a CTL, $\underline{\mathbf{M}}$ is given by

$$\underline{\mathbf{M}} = \begin{bmatrix} \underline{\mathbf{0}} & -j\underline{\mathbf{Z}} \\ -j\underline{\mathbf{Y}} & \underline{\mathbf{0}} \end{bmatrix} \quad (8)$$

and $\underline{\mathbf{0}}$ is the 2×2 zero matrix. In the absence of gain and loss, the $\underline{\mathbf{M}}$ matrix satisfies the J-Hermiticity property [4] (also see Ch. 6 in [44] or Ch. 3 in [45]), i.e.,

$$\underline{\mathbf{M}}^\dagger = \underline{\mathbf{J}}\,\underline{\mathbf{M}}\,\underline{\mathbf{J}}^{-1}, \quad \underline{\mathbf{J}} = \underline{\mathbf{J}}^\dagger = \underline{\mathbf{J}}^{-1} = \begin{bmatrix} \underline{\mathbf{0}} & \underline{\mathbf{1}} \\ \underline{\mathbf{1}} & \underline{\mathbf{0}} \end{bmatrix} \quad (9)$$

where the dagger symbol † denotes complex conjugate transpose and $\underline{\mathbf{1}}$ is 2×2 identity matrix. We look for solution of (7) of the type $\mathbf{\Psi}(z) \propto e^{-jkz}$, that is hence rewritten as $-jk\mathbf{\Psi}(z) = -j\underline{\mathbf{M}}\mathbf{\Psi}(z)$. Hence, eigenmodes supported by the uniform CTL described by (7) are found by solving eigenvalue problem,

$$\underline{\mathbf{M}}\mathbf{\Psi}(z) = k\mathbf{\Psi}(z) \quad (10)$$

where $k$ is a wavenumber of the CTL and here it represents the eigenvalue of the system. When the left equation of (9) holds with a matrix $\underline{\mathbf{J}}$ that is Hermitian and unitary, the eigenvalues (wavenumbers $k$) of $\underline{\mathbf{M}}$ are real as long as there are no gain and loss. In Appendix B we also prove, in a different way, that there is no degeneracy, i.e., $\underline{\mathbf{M}}$ is diagonalizable and has real eigenvalues, in the absence of gain and loss. We remind that in this paper the "gain and loss balance condition" is defined as the condition that guarantees the existence of an EPD in such CTL system described in (7). In the next section we will explore balanced gain and loss and assess the lesser strict condition of $\mathcal{PT}$-symmetry on the matrix $\underline{\mathbf{M}}$ such that an EPD and real $k$ eigenvalues could be found.

When $\underline{\mathbf{M}}$ is diagonalizable (for example, when $\underline{\mathbf{M}}$ is Hermitian), one can construct a *similarity transformation* of $\underline{\mathbf{M}}$ into a diagonal matrix $\underline{\mathbf{\Lambda}}$ containing all the eigenvalues $k$ as

$$\underline{\mathbf{M}} = \underline{\mathbf{U}}\,\underline{\mathbf{\Lambda}}\,\underline{\mathbf{U}}^{-1} \quad (11)$$

where $\underline{\mathbf{U}}$ is a non-singular similarity transformation, and its column are the four regular eigenvectors $\mathbf{\Psi}_n$ of $\underline{\mathbf{M}}$ namely $\underline{\mathbf{U}} = \begin{bmatrix} \mathbf{\Psi}_1 \mid \mathbf{\Psi}_2 \mid \mathbf{\Psi}_3 \mid \mathbf{\Psi}_4 \end{bmatrix}$. Note that such $\underline{\mathbf{U}}$ becomes singular when $\underline{\mathbf{M}}$ is non-diagonalizable (at the EPD) because at least two eigenvectors coalesce and hence they are no longer independent.

Since $\underline{\mathbf{M}}$ is a 4×4 matrix, (10) has four $k$-solutions obeying symmetry property in reciprocal systems, i.e., $\pm k$ are both solutions. Furthermore, we recall that in a lossless CTL, $k$ and $k^*$ are both solutions. Because of the structure of $\underline{\mathbf{M}}$, equation (10) is reduced to two simpler eigenvalue problems of two dimensions when only eigenvalues are requested. In uniform CTLs, these eigenvalue equations are readily found as

$$-\underline{\underline{\mathbf{ZY}}}\mathbf{V}(z) = k^2\mathbf{V}(z), \quad -\underline{\underline{\mathbf{YZ}}}\mathbf{I}(z) = k^2\mathbf{I}(z). \quad (12)$$

Each one provides the four eigenvalues. It is straightforward to see that in the absence of gain and loss each of the characteristic matrices $\underline{\underline{\mathbf{ZY}}}$ or $\underline{\underline{\mathbf{YZ}}}$ is *Hermitian* [43] and accordingly a lossless uniform CTL possesses entirely real-$k$ eigenvalues. A trivial case when the characteristics matrix $\underline{\underline{\mathbf{ZY}}}$ or $\underline{\underline{\mathbf{YZ}}}$ is non-Hermitian occurs at a cutoff condition, at which the matrices $\underline{\mathbf{Z}}$ and/or $\underline{\mathbf{Y}}$ become no-longer positive definite as the case of TLs describing TE/TM modes in rectangular waveguides and their cutoff for instance; see Ch. 8 in [46]. Here we do not investigate cutoff–related degeneracy but rather those special ones occurring when gain and loss balance is achieved; i.e., when $\underline{\underline{\mathbf{ZY}}}$ or $\underline{\underline{\mathbf{YZ}}}$ are not Hermitian at any considered frequency, as described in Section IV. In general, impedance and admittance matrices do not commute, i.e., $\underline{\underline{\mathbf{ZY}}} \neq \underline{\underline{\mathbf{YZ}}}$. A necessary and sufficient condition for these two matrices to commute is that both $\underline{\mathbf{Z}}$ and $\underline{\mathbf{Y}}$ share all of their eigenvectors, i.e., they are simultaneously diagonalizable [37], [40]. This is obtainable in several kinds of waveguide structures. For example, in lossless multi-conductor transmission lines in a homogenous environment [42], [43], the product of $\underline{\underline{\mathbf{ZY}}}$ is given by $\underline{\underline{\mathbf{ZY}}} = -\omega^2 \varepsilon \mu \underline{\mathbf{1}}$; and thus they commute (see Ch. 5 in [42]). The commutation property is important when examining the characteristics impedance of the system (not investigated here for the sake of conciseness). However, in general $\underline{\mathbf{Z}}$ and $\underline{\mathbf{Y}}$ might not commute and in the following we consider this more general case (without resorting to any particular assumption rather than they are symmetric; and $\underline{\mathbf{L}}$ and $\underline{\mathbf{C}}$ are also positive definite while $\underline{\mathbf{R}}$ and $\underline{\mathbf{G}}$ are symmetric and represent losses and/or gain) as will be further discussed in Section IV.

*B. Transfer matrix*

Solution of (7), with a certain boundary condition $\mathbf{\Psi}(z_0) = \mathbf{\Psi}_0$ at a certain coordinate $z_0$ inside a uniform CTL segment, is found by representing the state vector solution at a coordinate $z_1$ using

$$\mathbf{\Psi}(z_1) = \underline{\mathbf{T}}(z_1, z_0)\mathbf{\Psi}(z_0) \quad (13)$$

where we define $\underline{\mathbf{T}}(z_1, z_0)$ as the *transfer matrix* which translates the state vector $\mathbf{\Psi}(z)$ between two points $z_0$ and $z_1$ along the $z$ axis. Within a uniform segment of a CTL the transfer matrix is easily calculated as

$$\underline{\mathbf{T}}(z_1, z_0) = \exp[-j(z_1 - z_0)\underline{\mathbf{M}}] \qquad (14)$$

and the transfer matrix satisfies the group property $\underline{\mathbf{T}}(z_2, z_0) = \underline{\mathbf{T}}(z_2, z_1)\underline{\mathbf{T}}(z_1, z_0)$ as well as the J-unitarity (see Ch. 6 and 9 in [44]), i.e.,

$$\underline{\mathbf{T}}^{\dagger}(z_1, z_0) = \underline{\mathbf{J}}\,\underline{\mathbf{T}}^{-1}(z_1, z_0)\,\underline{\mathbf{J}} \qquad (15)$$

which is a consequence of reciprocity restriction [4] and $\underline{\mathbf{J}}$ is defined in (9), in addition to the symmetry property $\underline{\mathbf{T}}(z_1, z_0)\underline{\mathbf{T}}(z_0, z_1) = \underline{\mathbf{1}}$, where $\underline{\mathbf{1}}$ is the 4×4 identity matrix.

### C. Exceptional points of degeneracy and transfer matrix

We consider second order degeneracies that satisfy the wavenumber symmetry property, associated with the algebraic multiplicity of the eigenvalues or the wavenumbers denoted by $m=2$. Therefore, the system matrix $\underline{\mathbf{M}}$ in (8) cannot be represented as in (11), and it is rather similar to a Jordan form

$$\underline{\mathbf{M}} = \underline{\mathbf{W}} \begin{bmatrix} \underline{\mathbf{\Lambda}}^+ & \underline{\mathbf{0}} \\ \underline{\mathbf{0}} & \underline{\mathbf{\Lambda}}^- \end{bmatrix} \underline{\mathbf{W}}^{-1},$$
$$\underline{\mathbf{\Lambda}}^+ = \begin{pmatrix} k_e & 1 \\ 0 & k_e \end{pmatrix}, \quad \underline{\mathbf{\Lambda}}^- = \begin{pmatrix} -k_e & 1 \\ 0 & -k_e \end{pmatrix} \qquad (16)$$

implying a second order degeneracy (see Ch. 7 in [48]). Here, the 4×4 matrix $\underline{\mathbf{W}}$ is a non-singular similarity transformation whose columns are two regular eigenvectors and two generalized eigenvectors corresponding to wavenumber solutions $k_e$ and $-k_e$, each with multiplicity of two, and $\underline{\mathbf{\Lambda}}^+$, and $\underline{\mathbf{\Lambda}}^-$ in (16) are 2×2 Jordan blocks, i.e., they are non-diagonalizable. As such, matrix $\underline{\mathbf{W}}$ is written as $\underline{\mathbf{W}} = \begin{bmatrix} \mathbf{\Psi}_1 | \mathbf{\Psi}_1^g | \mathbf{\Psi}_2 | \mathbf{\Psi}_2^g \end{bmatrix}$ (see Ch. 7.8 in [48]) where $\mathbf{\Psi}_{1,2}$ are the two regular eigenvectors associated with the eigenvalues $\pm k_e$, respectively, and $\mathbf{\Psi}_{1,2}^g$ are two generalized eigenvectors that are given by $\mathbf{\Psi}_{1,2} = (\underline{\mathbf{M}} \mp k_e \underline{\mathbf{1}})\mathbf{\Psi}_{1,2}^g$ (see Ch. 7.8 in [48]). A symbolic evolution for one pair of eigenvectors near a second order EPD is depicted in Fig. 2 showing they coalesce when $\omega = \omega_e$.

The characteristic equations for the eigenvalues of $\underline{\mathbf{M}}$ in (10) can be cast in a simple scalar polynomial of order four in $k$, whose solutions provide the four eigenvalues (i.e., the wavenumbers). Also from (12) one can write the characteristic polynomial for the eigenvalues of a 2×2 matrix (see p. 14-1 in [47]) as

$$k^4 + k^2 \,\mathrm{Tr}\left(\underline{\underline{\mathbf{ZY}}}\right) - \det\left(\underline{\underline{\mathbf{ZY}}}\right) = 0 \qquad (17)$$

where Tr and det are the trace and the determinant, respectively, and the analytic expression of eigenvalues are reported in Appendix B for the special lossless case. In the next two sections we discuss wavenumbers and conditions for EPD occurrence in gain and loss balanced CTL for both the symmetric ($\mathcal{PT}$-symmetry) and the asymmetric cases.

### IV. SECOND ORDER EPD IN UNIFORM SYMMETRIC CTLS

The examples in Fig. 1(a) and (b) can be well represented by those corresponding cases in Fig. 3 (a) and (b) respectively. These two cases represent uniform or quasi-uniform waveguides (meaning the structure is periodic but the period is subwavelength such that the coupled waveguide structure can be considered uniform in $z$, as the case of loading a CTL with densely distributed amplifiers). Results in this section pertain to a uniform symmetric guiding system made of two grounded microstrips, whose physical parameters are given in Appendix A as well as its equivalent CTL parameters. The other example of topologically asymmetric CTL will be provided in Section V.

#### A. Conditions on the CTL parameters and $\mathcal{PT}$-symmetry

The symmetric two-microstrip coupled lines are described by a CTL with symmetric and positive definite, per unit length capacitance and inductance matrices (5). In absence of loss and gain the system does not develop any EPD except at zero frequency which is a trivial condition (the proof of this statement is detailed in Appendix B). An EPD at a non-zero radian frequency, denoted by $\omega_e$, is obtained when gain and losses are introduced, represented here by negative/positive per-unit-length resistance and/or conductance. We show here the conditions under which the CTL system matrix $\underline{\mathbf{M}}$ can be written in the form (16), thus exhibiting a degenerate mode at $\omega = \omega_e$.

For simplicity we assume that the per-unit-length resistance and conductance matrices are diagonal. In this case, the $\mathcal{PT}$-symmetry condition is equivalent to $\underline{\underline{\mathbf{Z}}} = -\underline{\underline{\mathbf{\Gamma}}}\,\underline{\underline{\mathbf{Z}}}^{\dagger}\,\underline{\underline{\mathbf{\Gamma}}}$, $\underline{\underline{\mathbf{Y}}} = -\underline{\underline{\mathbf{\Gamma}}}\,\underline{\underline{\mathbf{Y}}}^{\dagger}\,\underline{\underline{\mathbf{\Gamma}}}$ when applied to CTLs with gain and loss, and $\underline{\underline{\mathbf{\Gamma}}}$ is the 2×2 backward or reverse identity matrix, i.e., with unity anti-diagonal elements. This symmetric condition is the analog of refractive index conjugate symmetry often utilized in $\mathcal{PT}$-symmetric optics [17], [24].

In this section we consider losses that are introduced as per-unit-length series resistance $R_2$ in TL2 while gain is introduced as per-unit-length negative series resistance $R_1$ in TL1. Therefore, in this example the series impedance per unit length matrix $\underline{\underline{\mathbf{R}}}$ is diagonal whereas $\underline{\underline{\mathbf{G}}} = \underline{\underline{\mathbf{0}}}$. Accordingly, in this case the matrices $\underline{\underline{\mathbf{ZY}}}$ and $\underline{\underline{\mathbf{YZ}}}$ in (12) are non-Hermitian. Therefore, we consider the gain and loss topological symmetry condition

$$R_2 = -R_1 = R. \qquad (18)$$

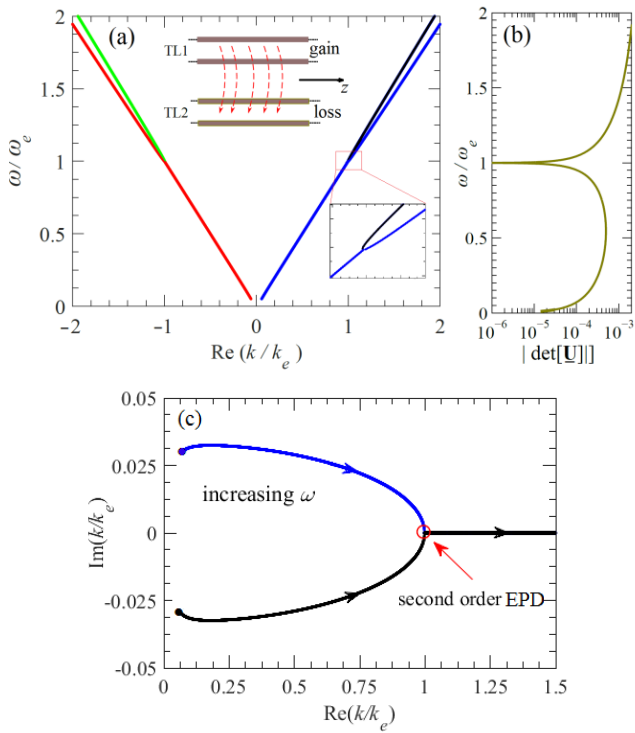

Fig. 4. (a) Dispersion diagram of the wavenumber $k$ versus angular frequency for a 2-TL with balanced gain-loss; developing an exceptional point (at $\omega=\omega_e$, $k=k_e$) at which bifurcation of two complex modes into two real modes takes place. (b) The magnitude of determinant of the similarity transformation $\underline{\mathbf{U}}$ in (11). The CTL parameters are given in Appendix A. (c) Complex $k$ space near a second order EPD. Trajectory of complex $k$ in the Re($k$)-Im($k$) plane varying as a function of $\omega$. Modes have complex wavenumber for $\omega < \omega_e$ (only the real part is shown) and purely real wavenumbers for $\omega > \omega_e$. PT-symmetry occurs for $\omega = \omega_e$, $k = k_e$ where only real modes are observed. (c) The positive branch of the wavenumber evaluated at varying as a function of frequency.

in which we denote $R$ as the distributed gain and loss parameter (real and positive number).

For the CTL under consideration with gain and loss, for simplicity we further assume the following symmetries

$$C_{11} = C_{22} = C, \quad L_{11} = L_{22} = L,$$
$$L_{12} = L_{21} = L_m, \quad C_{12} = C_{21} = C_m. \quad (19)$$

Moreover, the positive definite condition of the inductance and capacitance matrices implies $L > L_m$ and $C > C_m$. When both (18) and (19) are realized simultaneously for non-zero gain and loss parameter $R$, we refer to the CTL as being $\mathcal{PT}$-symmetric. In order that (17) satisfies the second order degeneracy condition at $\omega = \omega_e$, i.e., $\left(k^2 - k_e^2\right)^2 = 0$, where $\pm k_e$ are the degenerate state eigenvalues, each of multiplicity two, the following conditions must be verified

$$\mathrm{Tr}\left(\underline{\underline{\mathbf{ZY}}}\right) = -2k_e^2, \quad \text{and} \quad \det\left(\underline{\underline{\mathbf{ZY}}}\right) = -k_e^4. \quad (20)$$

However, the coincidence of two eigenvalues is not sufficient for the very special EPD condition examined in this paper, we also need *pairs of eigenvectors* to coalesce at radian frequency $\omega_e$, associated to each of the two wavenumbers $k_e$ and $-k_e$. as schematically represented in Fig. 2. In other words, if at $\omega_e$ we have a degeneracy condition at $k_e$ of order two, we also have another one at $-k_e$ of order two. Therefore, we need to be sure that some other necessary condition is also simultaneously satisfied. In particular, we impose that the Jordan block similarity restriction (16) is imposed at $\omega_e$ and $\pm k_e$ leading to an additional condition (along with those in (18) through (20)) that reads

$$\left[C^2L^2 + C_m^2L_m^2 + \left(C^2 - C_m^2\right)\left(\frac{R}{\omega_e}\right)^2 - \left(C^2L_m^2 + C_m^2L^2\right)\right] = \left(k_e/\omega_e\right)^4,$$
$$\omega_e^2\left[LC - L_mC_m\right] = k_e^2 \quad (21)$$

These conditions, (20) and (21), together, are necessary and sufficient to realize a second order EPD, under the simplifying geometrical symmetry assumptions in (18) and (19). By properly choosing the parameters $L$, $C$, and $L_m$, and a certain $C_m = C_{m,e}$, we obtain the value of $R$ that provides for an EPD by imposing (20) and (21) at a desired radian frequency $\omega_e$ and wavenumber $k_e$.

### B. Second order EPD and $\mathcal{PT}$-symmetry

To demonstrate the existence of an EPD in a CTL, we show in Fig. 4(a) the dispersion relation of the modes supported by the CTL system designed to exhibit an EPD at radian frequency $\omega_e$ (i.e., at frequency $f_e = \omega_e/(2\pi) = 1$ GHz) and wavenumber $k_e$. The values of the CTL parameters are listed in the Appendix A, corresponding to lossy coupled microstrip lines with distributed amplifiers modeled with distributed negative resistance. The resulting eigenvalues at the EPD are at both $k = k_e$ and $k = -k_e$, where from (21) $k_e = \omega_e\sqrt{LC - L_mC_m} = 28.66$ m$^{-1}$. Note that the dispersion diagram supports complex modes in general (modes with complex wavenumber $k$ for real frequencies), but in Fig. 4(a) we plot only the real part of the wavenumber to show the EPD, at which a bifurcation of modes occurs.

To illustrate that the system has indeed developed an EPD at $\omega = \omega_e$, we utilize the similarity transformation in (11) that is valid for $\omega \neq \omega_e$ and calculate its determinant, namely $|\det[\underline{\mathbf{U}}]|$, varying as a function of frequency. The $|\det[\underline{\mathbf{U}}]|$ measure is used here identify the occurrence of any EPD in the system, though it cannot necessarily recover the order of such degeneracy. Indeed, the 4×4 matrix $\underline{\mathbf{U}}$ contains four independent eigenvectors as columns for any $\omega \neq \omega_e$, but $|\det[\underline{\mathbf{U}}]| = 0$ when at least two eigenvectors coalesce, hence $\underline{\mathbf{U}}$ becomes singular. This implies also that a diagonal form of $\underline{\mathbf{M}}$ cannot be obtained at $\omega = \omega_e$, and only at $\omega = \omega_e$, $\underline{\mathbf{M}}$ becomes similar to a matrix that contains Jordan blocks as in (16). (We recall that diagonalization of the matrix $\underline{\mathbf{M}}$ is a sufficient condition for the existence of four independent eigenvectors, see p. 4-7 in [47].) Interestingly, the measure $|\det[\mathbf{U}]|$ is proportional to the solid angle between the four complex vectors in a four dimensional space [49], and was utilized in [50] to detect second order EPDs. This feature is shown in Fig. 4(b) where it is seen that at $\omega = \omega_e$ one has $|\det[\mathbf{U}]| \to 0$ as an indication that the system is undergoing a degeneracy (of a second order in this case).

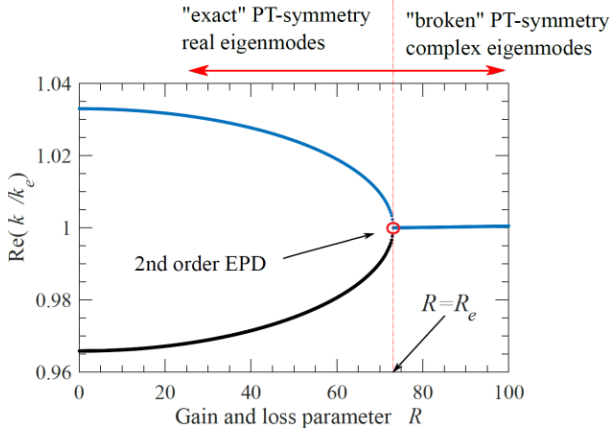

Fig. 5. The positive branch of the wavenumber evaluated at $\omega = \omega_e$ varying as a function of the gain and loss parameter $R$. When $R$ reaches a critical value, an EPD is manifested.

In Fig. 4(c), we analyzing the complex Re($k$)–Im($k$) wavenumber. There the trajectory of the mode's wavenumber $k$ for two modes with $k(\omega)$ and $k^*(\omega)$, both with positive real part, i.e., with phase propagating along the +$z$-direction, is plotted with increasing angular frequency from $0.5\omega_e$ to $1.5\omega_e$. The second order EPD occurs at $\omega_e$, at which the two blue and black curves meet. For frequencies such that $\omega > \omega_e$ the two wavenumbers are purely real, despite the presence of losses and gain.

In particular, for the radian frequency lower than the EPD's $\omega_e$, modes are complex with complex conjugate wavenumbers $k(\omega), k^*(\omega), -k(\omega), -k^*(\omega)$. Note that $k(\omega)$ and $k^*(\omega)$ are both solution in a lossless system, here instead with *both* losses and gain such symmetry is in principle not necessarily expected. Indeed, for $\omega < \omega_e$ the right branch of the dispersion diagram, with positive Re($k$), comprises two modes, with $k(\omega)$ and $k^*(\omega)$: one exhibits exponential growth in the positive $z$-direction while the other decays exponentially in the positive $z$-direction. The other branch with negative Re($k$), represents also two modes with phase propagation in the –$z$-direction that behave analogously. In this frequency range ($\omega < \omega_e$) modes have complex wavenumbers despite perfect and symmetrical gain-loss impedance balance. At $\omega_e$ pairs of eigenvalues (wavenumbers) in each branch coalesce at $\pm k_e$, hence assuming a vanishing imaginary part and forming two

EPDs, for each of $\pm k_e$. For $\omega > \omega_e$ modes split into pairs with purely *real wavenumber*, hence they do not exhibit neither exponential growth nor decay, despite the presence of loss and gain.

According to the symmetries assumed in the CTL and governed by (18) and (19), it is easy to show that the matrix $\mathbf{M}$ satisfies the relation $\mathbf{M}^\dagger = \mathbf{\Gamma}\,\mathbf{M}\,\mathbf{\Gamma}^{-1}$ where here $\mathbf{\Gamma}$ is the backward, also called reverse, identity matrix, i.e., the 4×4 matrix having ones on the main anti-diagonal.

We recall that every complex system with a real spectrum is *pseudo-Hermitian* [51], i.e., the pseudo-Hermitian system matrix can be written as $\mathbf{M}^\dagger = \mathbf{\eta}\,\mathbf{M}\,\mathbf{\eta}^{-1}$ where $\mathbf{\eta}$ is a Hermitian and unitary transformation matrix (additionality, if $\mathbf{\eta}$ is also positive-definite, then $\mathbf{M}$ has entirely real spectra [51]–[53]). Indeed, as shown by Mostafazadeh [51]–[54] all the $\mathcal{PT}$-symmetric Hamiltonians studied in the literature exhibited such property. Note that as described above, we have $\mathbf{M}^\dagger = \mathbf{\Gamma}\,\mathbf{M}\,\mathbf{\Gamma}^{-1}$, where $\mathbf{\Gamma}$ is indeed unitary and Hermitian, therefore the system matrix $\mathbf{M}$ in (10) is pseudo Hermitian. On one hand, a pseudo-Hermitian matrix possesses either purely real spectrum or complex conjugates pairs of eigenvalues [51], which is indeed what is depicted in Fig. 4 depending on a system parameter. One the other hand, $\mathcal{PT}$-symmetry strictly is not a necessary nor sufficient condition for a system to develop entirely real spectrum [51]–[54].

To realize such an EPD at a given $\omega_e$ and $k_e$, a specific amount of gain and loss must be satisfied, as shown in Fig. 5 where the gain and loss parameter $R$ is varied at the fixed frequency $\omega = \omega_e$. Only the positive Re($k$) branch is shown for simplicity. Two modes coalesce at a critical value of the parameter $R$, denoted by $R_e$. For $R < R_e$ the two modes are distinct and have purely real wavenumbers. The condition $R = R_e$ correspond to the occurrence of the EPD at $\omega = \omega_e$ and it also designates the onset of *"$\mathcal{PT}$-symmetry breaking"* [17], [55].

*C. Analysis of EPD perturbation*

It is important to point out that near the second order EPD (near in frequency or in some other parameter), the two eigenvalues of the CTL (i.e., those with positive wavenumbers $k$) can be written as a small perturbation of that relative to the degeneracy condition $k = k_e$, in terms of a fractional power expansion as

$$k_n(\omega) \cong k_e + a_n \delta^{1/2} + b_n \delta + ... \quad (22)$$

Analogous discussion is valid near the EPD wavenumber at $k = -k_e$. Here $a_n$ and $b_n$ are the fractional series expansion coefficients for the two modes in the +$z$-direction, denoted by $n=1,2$, and $\delta$ is the small perturbation parameter of the system about the EPD (refer to [56] where the parameters $a$ and $b$ in (21) can be obtained from the system's exact dispersion relation). In this example, one can assumes $\delta \equiv (\omega - \omega_e)$ representing the frequency detuning away from the ideal degeneracy frequency and therefore (22) represents the dispersion relation [the principle square root for $\delta$ is taken in (22)]. This fractional power expansion, called Puiseux series, is a direct consequence of the Jordan Block similarity [4], [56], [57]. The branches of dispersion relation in Fig. 4(c) are well fitted by (22) in the neighborhood of $\omega = \omega_e$.

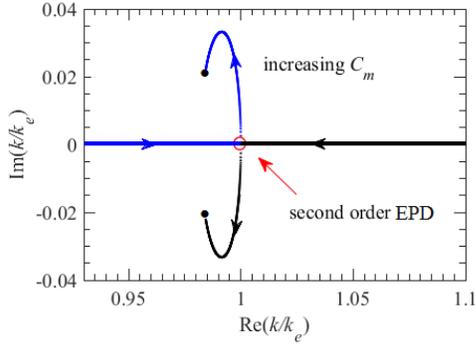

Fig. 6. Complex $k$ space near a second order EPD. Trajectory of complex $k$ in the Re($k$)-Im($k$) plane varying as a function of $C_m$, for $\omega=\omega_e$, where the EPD is manifested at $C_m = C_{m,e}$. Note that for symmetry reasons, we have omitted the –Re($k$) branches.

To further elaborate on different perturbation effects in CTL, we inspect how the detuning of some CTL parameter, away from the proper value for which EPD is realized, modifies the dispersion diagram. In Fig. 6, we plot the modal wavenumbers in the complex Re($k$)-Im($k$) plane varying the mutual per-unit-length capacitance $C_m$, at the angular frequency $\omega = \omega_e$. We denote now as $C_{m,e}$ the nominal coupling capacitance $C_m$ per-unit-length that has been used to achieve an EPD for given parameters as in Appendix A, except for $C_m$ that is varying. The trajectory of the complex $k$ is plotted versus $C_m$ varying from $0.4 C_{m,e}$ to $1.2 C_{m,e}$. For $C_m > C_{m,e}$ the wavenumbers are complex while for $C_m < C_{m,e}$ the wavenumbers are purely real. This wavenumber behavior can be easily explained by assuming $\delta \equiv C_m - C_{m,e}$ in (22) which provides also the curvature of the $k$-trajectories in the complex $k$-plane in proximity of the EPD. Analogous effects are observed when detuning some other structural parameter.

The trajectory of the real part of the modal wavenumber $k$, varying as the coupling capacitance $C_m$ changes is shown in Fig. 7(a), assuming three different gain and loss $R$ values [see (18)] at the angular frequency $\omega_e = 2\pi 10^9$ rad/s for the three cases, where all CTL values are provided in Appendix A. EPD wavenumbers are all normalized with respect to the wavenumber $k_e = \omega_e \sqrt{LC - L_m C_{m,e}} = 28.66$ m$^{-1}$; which is the EPD wavenumber relative to the case with $R$=73 Ω/m. Interestingly, the system develops EPDs at two values of $C_m$ once the gain and loss parameters $R$ has been fixed: such EPDs are denoted by black dots in Fig. 7(a). In all curves, the wavenumber $k$ is normalized to the nominal $k_e$ value just provided, obtained with the parameters in Appendix A. Since EPD is achieved for not only one value of mutual capacitance, but two, this indicates that there is some freedom in the choice of gain and loss parameter $R$. For example, for the case with $R = 73$ Ω/m (as in Appendix A), an EPD can be found when the coupling capacitance is selected to be either $C_m/C \approx 0.21$, corresponding to the case of $C_m = C_{m,e}$ with $k=k_e$ described above, or $C_m/C \approx 0.38$ that occurs at $k=0.98 k_e$ (recall that $C$ is the CTL per-unit-length self-capacitance in (19) and given in appendix A). However, a very large gain and loss parameter $R$ may not permit in practice the structure to develop an EPD, since such requirement would render the CTL parameters (per unit length capacitance for instance) not realizable with standard passive components. This effect of detuning implies that other structural parameters or frequency of the EPD could be adjusted accordingly for an EPD to manifest. In other words, these results show some flexibility in the possibility to achieve an EPD. For the case shown in Fig. 7(a) with $R = 1$ Ω/m, the EPD can be only achieved for two values of the coupling capacitance $C_m$ that are almost identical implying that the flexibility in designing a second order EPD in *uniform* CTLs with *small values of the gain and loss* parameter $R$ is no longer guaranteed. Furthermore, for small values of $R$, the resulting mode trajectories in Fig. 6(a) do not exhibit large radius of curvature near the EPD as for the case of larger values of $R$. Obviously, in the limiting case when $R = 0$ Ω/m (i.e., no loss and no gain) the system loses the capability to exhibit an EPD, because the $\underline{\underline{ZY}}$ matrix in (12) becomes Hermitian in this case and hence it could be diagonalized as shown in Appendix A.

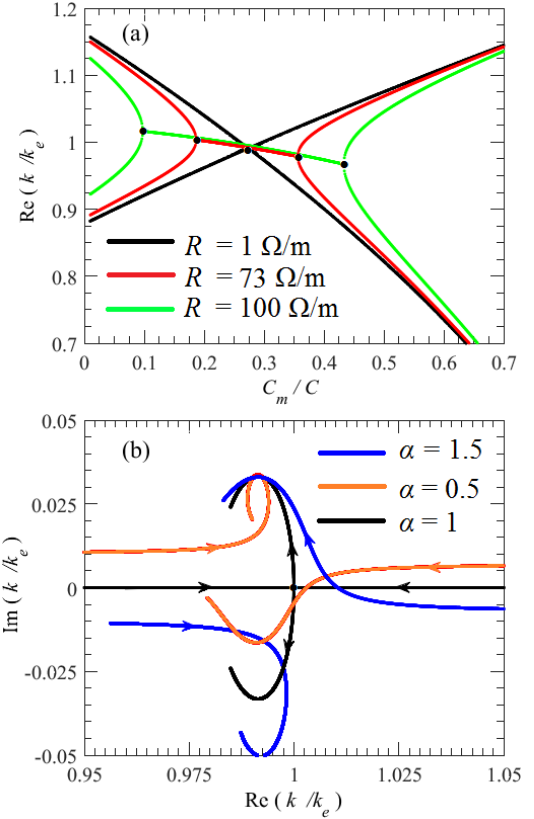

Fig. 7. Real part of the complex wavenumber varying as a function of the coupling distributed capacitance $C_m$ calculated at $\omega = \omega_e$. The black dots indicate the EPD. (b) Real part of the complex wavenumber plane varying as a function of the gain/loss parameter imbalance $\alpha$. $k_e$ is the same as that in Fig. 4.

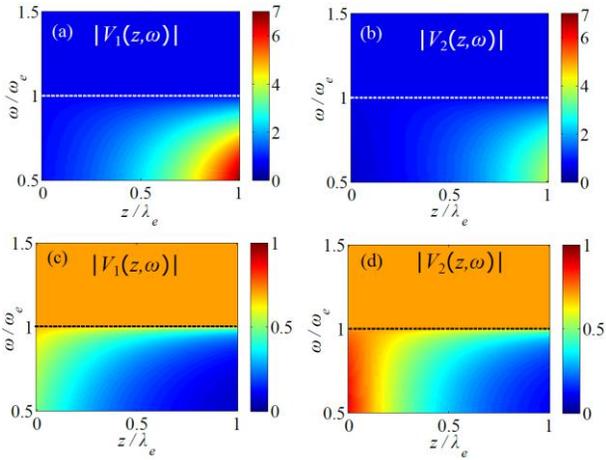

Fig. 8. Behavior of normalized voltages $V_1(z)$ and $V_2(z)$ of the first mode in (a) and (b) and second mode in (d) and (d) in a 2-TL system (parameters in Appendix A). In (a) and (b) we plot of the voltages corresponding to the mode exhibiting exponential growth in the +z-direction, for $\omega < \omega_e$ (the mode with blue color in Fig. 4(a)). The same mode has real wavenumber and also $V_1(z) = V_2(z)$ for $\omega > \omega_e$. In (c) and (d) we plot voltages corresponding to the mode exhibiting exponential decaying in the +z-direction for $\omega < \omega_e$ (the mode with black color in Fig. 4). The same mode has real wavenumber and also $V_1(z) = V_2(z)$ for $\omega > \omega_e$. Note that for each mode voltage and current distributions in TL1 and TL2 are not symmetric for $\omega < \omega_e$, whereas they are exactly identically for $\omega > \omega_e$, following the PT-symmetry properties.

Another insightful investigation shows the effect of *imbalance or asymmetry* of loss and gain by assuming gain and loss do not satisfy (18) i.e., they differ by a factor $\alpha$. Specifically, here we assume loss $R_2 = R$ in TL2 and gain $R_1 = -\alpha R$ in TL2, while keeping all other CTL parameters fixed based on (19) and the values in Appendix A. Therefore, the imperfect gain-loss symmetry breaks the $\mathcal{PT}$-symmetry, hence real modes and EPDs can no longer be found unless other structural parameters are adjusted (see Section V). In Fig. 7(b) the trajectory of the complex wavenumber $k$ is shown for different values of $\alpha$, for growing coupling capacitance $C_m$ in the range $0.5 C_{m,e} < C_m < 3.3 C_{m,e}$ in the direction of the arrows. We see that the bifurcation of modes occurring when $\alpha=1$ (the perfect loss/gain symmetry case) is lost for the other two cases as there is gain-loss asymmetry. Nonetheless, when these balance imperfections are small (treated as small perturbations) we still can benefit from the properties associated with the EPD since the wavenumber $k$ may not differ substantially from $k_e$. In Section V however we will show that EPD can occur also in this asymmetric case when other parameters than $R$ are also changed.

### D. Voltages and currents of the CTL near the EPD

The three different regimes ($\omega < \omega_e$, $\omega = \omega_e$ and $\omega > \omega_e$) exhibit distinct characteristics of the voltage $z$-dependency in the CTL. For $\omega < \omega_e$, based on Fig. 4(c), the two modes with $k(\omega)$ and $k^*(\omega)$ grow and decay along the +z-direction since they have complex and conjugate wavenumbers (we assume the CTL parameters listed in Appendix A). In order to show voltages, we calculate the normalized eigenvectors $\Psi_n(z)$ of the system in (10) relative to the *n*-modes propagating in the +z-direction, starting from a given arbitrary value $\Psi_n(0)$. Here $n$ = 1,2,3,4, however only two modes are shown in Figs. 4-8 because of the $\pm k$ symmetry of modal wavenumbers in reciprocal systems. In Fig. 8(a,b) we show the normalized state vector's voltages $V_1(z)$ and $V_1(z)$ of the CTL mode exhibiting exponential growth (and positive Re($k$)) in both TLs in the +z-direction for $\omega < \omega_e$, in accord to the blue curve in Fig. 4(c). The state vector's voltage normalization is done such that the state eigenvectors have a norm $\|\Psi_n(0)\| = 1$, and the $z$ coordinate is normalized by the wavelength $\lambda_e = 2\pi / k_e$. Plots are shown in a linear color scale. The evolution of voltages in TL1 and TL2 is not symmetric, i.e., $V_1(z) \neq V_2(z)$ for $\omega < \omega_e$ however $V_1(z) = V_2(z)$ for $\omega > \omega_e$. The other CTL mode shown in Fig. 8(c,d), relative to the black curve in Fig. 4(c), exhibits exponential decay (and positive Re($k$)) for $\omega < \omega_e$. The asymmetry between $V_1(z)$ and $V_2(z)$ [for each of the two modes with decay and exponential growth for $\omega < \omega_e$, represented by blue and black curves in Fig. 4(c)], is evident comparing (a) with (b) and (c) with (d). Indeed, TL1 has losses and TL2 has balanced gain, and this allows for asymmetric transmission and resonance characteristics highlighted in [28], [58] that should be investigated in depth in the future. At $\omega = \omega_e$, solutions for the state vector in (7) are degenerate and given in terms of two propagating modes as $\Psi_{1,2}(z) \propto \Psi(0) e^{\pm j k_e z}$, and two other modes that contain also diverging terms as $\Psi_{3,4}(z) \propto z \Psi(0) e^{\pm j k_e z}$. (Note the algebraic $z$-factor in front of $\Psi(0)$ for modes with $n$ = 3,4, that is the result from the EPD condition and the similarity to a Jordan block.) Therefore, exactly at $\omega = \omega_e$, the voltage in the TLs of the two degenerate modes in the +z-direction (i.e., for $n$ = 1 and 3, for example) would scale as $e^{-j k_e z}$ and $z e^{-j k_e z}$ (not shown in Fig. 8). This EPD characteristic results in a giant resonance in a CTL system of finite length, i.e., in a CTL resonator based on EPD modes. An analogous characteristic relative to giant resonances based on a fourth order degeneracy has been employed to conceive giant gain [7], [10] and new low-threshold electron beam-based oscillators in [59].

### V. SECOND ORDER EPD AND GAIN AND LOSS BALANCE IN UNIFORM ASYMMETRIC CTLS

The discussion in Section IV was devoted to coupled waveguides with topological symmetry, where loss and gain obey the conjugate symmetry condition (18). However, EPDs can be implemented in asymmetrical configurations of gain and loss in coupled waveguides [60]. In other words, $\mathcal{PT}$-symmetry is not a necessary condition to realize a second order EPD. We emphasize that such topological asymmetry is intriguing in terms of applications since exact symmetry is hard to achieve in practice. If an asymmetric gain and loss distribution allows an EPD we say we have an EPD enabled by the "gain and loss balance". Here we demonstrate the concept by considering an example of coupled waveguides such as the geometry of microstrip lines in Fig.1(b) and represented as in Fig. 3(b), where loss is present in both microstrips while gain is present in only one. This configuration is able to support an EPD. In particular, we assume, for practical purposes, that losses are represented in the CTL as a series per-unit length resistance in *both* TLs, $R_1 = R_2 = R_e = 73$ Ω/m thus accounting for metal losses. Gain is introduced as a shunt conductance $G$ (real and negative number), which represents the small signal distributed gain and could be realized via solid-state devices such as cross-coupled transistors. Here we define $-G$ as the positive gain parameter with units of S/m. Without further elucidating the details on how to realize the EPD in this case, we carry out the same analysis outlined in Section IV yet here for the asymmetric topology of gain and loss.

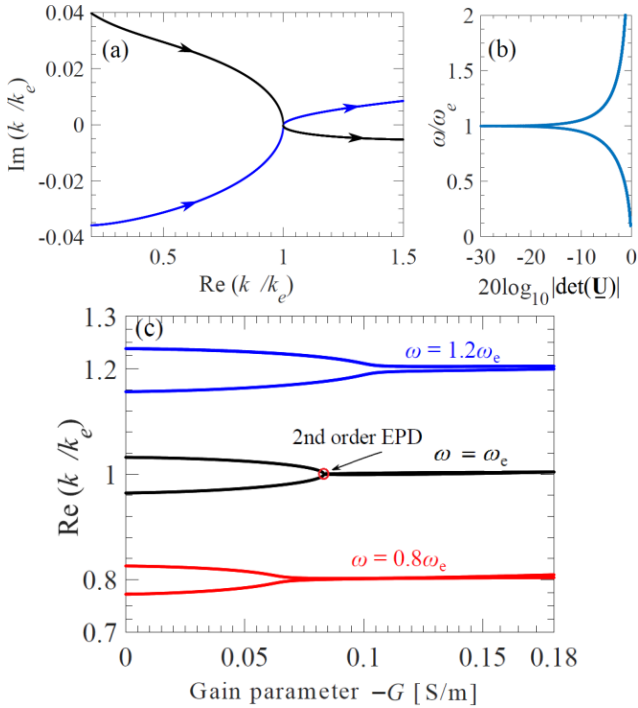

Fig. 9. Realization of a non-Hermitian EPD in uniform CTL with asymmetric gain and loss topology that is depicted in Fig. 3(b). (a) Trajectory of the normalized wavenumber in the complex $k$-plane varying with frequency. (b) The magnitude of the determinant of $\underline{\mathbf{U}}$. (c) Real part of the normalized wavenumber varying as a function of the shunt gain parameter $-G$ for three different frequencies. The exact EPD is realized when $G \approx -83.2$ mS/m and $\omega = \omega_e$.

We also allow some other CTL parameter to be asymmetric; contrary to what was assumed in Section IV. We aim at finding conditions on the CTL parameter $G$ as well as the second TL self-capacitance per-unit-length, denoted by $C_{22}$ in order for a second order EPD to manifest at the desired frequency $\omega_e = 2\pi 10^9$ rad/s. We then fix the rest of the CTL parameters as assumed in Section IV.B (the values of $R_c$, $L_{11}$, $L_{22}$, $L_m$, $C_{11}$ and $C_m$ that are given in Appendix A), and consequently we find the required values of $G$ and $C_{22}$ in order for an EPD to manifest. The value of $G$ that guarantees the existence of the EPD is $G \approx -83.2$ S/m while $C_{22} \approx 0.145$ nF/m. We point out that an EPD is realized with a minor asymmetry in the CTL intrinsic parameters (here $C_{22}/C_{11} \sim 1.18$) while the gain and loss topology demonstrates substantial topological asymmetry. The resulting values of $k_e$ at the EPD is complex, and $k_e$ is found to be $k_e = (28.2 - j0.13)\,\mathrm{m}^{-1}$. The relative behavior of the normalized complex wavenumber trajectory, namely Re($k/k_e$) varying as a function of frequency is depicted in Fig. 9(a) in which the trajectory exhibits a point of coalescence at $k = k_e$ and $\omega = \omega_e$. Only curves with the real part of the normalized wavenumber is plotted. The other two modes with Re($k$) < 0 are found immediately based on the $\pm k$ symmetry of modal wavenumbers in reciprocal systems. The existence of such second order EPD is proven by the vanishing determinant of the similarity matrix $\underline{\mathbf{U}}$ in (11) as seen from Fig. 9(b), since two eigenvectors coalesce at the EPD. Moreover, we note the qualitative differences between this scenario and what happens in topologically symmetric CTL, whose wavenumber trajectories are shown before in Fig. 4(c). In the topologically *asymmetric* case in Fig. 9(a), the bifurcation of modes results in two complex branching of the wavenumber for both $\omega < \omega_e$ and $\omega > \omega_e$. On the contrary the topological symmetry as in Section IV and in Fig. 4(c) shows a single bifurcation, i.e., both modes have purely real wavenumber for $\omega > \omega_e$.

We also show in Fig. 9(c) the real part of the normalized wavenumber $k$ varying as function of the gain parameter $-G$ for three different frequencies. Observe that the perfect EPD, enable by a balanced loss and gain, occurs at the designated condition when $\omega = \omega_e$. For frequencies in the vicinity of $\omega = \omega_e$, the Re($k/k_e$) plots show deformation from the ideal EPD condition in such a way that branches never coalesce for $\omega = 0.8\omega_e$ or $\omega = 1.2\omega_e$, for the considered CTL parameters. Moreover, the behavior of the wavenumbers for these two imperfect EPDs obey the perturbation formula (22) with $\delta$ representing either the frequency detuning from $\omega = \omega_e$ or the gain variation for the one that enables the EPD.

For practical viewpoint, it is natural to argue when such EPD does occur. Here it is sufficient to investigate the metric $|\det(\underline{\mathbf{U}})|$, because the vanishing of $|\det(\underline{\mathbf{U}})| = 0$ by itself is a necessary and sufficient condition for a *second order* EPD to occur. As such, the obtained EPD in Fig. 9(c) (black curve with $\omega = \omega_e$) has $|\det(\underline{\mathbf{U}})| = 0$ that identically vanishes when the gain parameter is equal to a critical value of $G = -83.2$ mS/m. Whereas for the other cases shown in Fig. 9(c), the minimum $|\det(\underline{\mathbf{U}})|$ varying as a function of the gain parameter, is 0.22 and 0.3 for $\omega = 0.8\omega_e$ and $\omega = 1.2\omega_e$ respectively. Note that higher order degeneracies, where more than two eigenvector coalesce as in [3], [4], [11], require a more elaborate figure of merit not discussed in this paper. The remarkable EPD condition due to gain and loss balance could be employed in designing amplifiers, distributed oscillators, radiating array oscillators, and sensors at microwave frequencies.

## VI. Conclusion

We have demonstrated a transmission line theory of guided waves in coupled waveguides with exceptional points of degeneracies (EPDs). We have shown two types of degeneracies that may occur in coupled waveguides, within the same unified theoretical framework, namely a second order EPD in balanced gain-loss uniform waveguides first satisfying $\mathcal{PT}$-symmetry (i.e., gain and loss topological symmetry) and also, importantly, a gain and loss balance condition without gain and loss symmetry. We have also developed a figure of merit measure for estimating the effect of imperfect coupling, unbalanced losses and topological symmetry breaking on the occurrence of EPDs.

The theoretical framework developed here applies to many structures operating from microwave to optical frequencies and can readily account for gain provided by the active devices. Because the EPD is associated with "slow-light" (electromagnetic wave with vanishing group velocity), extremely high Q-factor resonators can be designed as well high low-threshold nonlinear processes in the microwave region. A particular application of interest is the balance of gain and loss in high power traveling wave tubes, promising high efficiency when electron beams act as source of gain while distributed loads represent mechanism of losses or radiation losses. This could be done also in a pulse-compression scheme [9] where the electron beam acts also as a switch. The premise of EPDs would also benefit a large category of other applications, such as low-threshold microwave and terahertz sources, sensors, antennas, and RF circuits, including radiating array oscillators. Moreover, further investigation would be required toward experimentally demonstrating such conditions in coupled microstrip transmission lines, as an example.

## APPENDIX A: NUMERICAL PARAMETERS USED IN SIMULATIONS

The uniform (i.e., $z$-invariant) CTL used in this paper corresponds to two coupled microstrip lines, depicted in Fig. A1 and has the following parameters: strip width 3 mm, gap between strips 0.1 mm; substrate height 0.75 mm and dielectric constant of 2.2, with a ground plane underneath. The corresponding CTL per unit length parameters are $L_{11} = L_{22} = L = 0.1798\,\mu\text{H/m}$ for both TLs, the mutual inductance is $L_{12} = L_{21} = L_m = 49.235\,\text{nH/m}$ and both TLs have $C_{11} = C_{22} = C = 0.1227\,\text{nF/m}$, whereas the coupling capacitance $C_{m,e} = 25.83\,\text{pF/m}$ is chosen to develop a second order EPD at 1 GHz, when losses and gain parameters in (18) are equal to $R = 73\,\Omega/\text{m}$, i.e., when $R_{11} = -R_{22} = R$ and $R_{12} = R_{21} = 0$. The value of such distributed series loss resistance in one TL can be achieved using high loss metal traces in the microstrip (here Lead can be used whose electrical conductivity is $\sim 4.5 \times 10^6\,\text{S/m}$ providing per unit resistance in the microstrip of $\sim 73\,\Omega/\text{m}$ when the metal thickness is $\sim 10\,\mu\text{m}$), still being low-loss since $R < \omega L_{11}$. In the asymmetric case in Sec. V, we assume losses in both TLs such as $R_{11} = R_{22} = R$, $R_{12} = R_{21} = 0$ whereas gain is represented by the shunt per unit length negative conductance $G_{11} = -G$ with $G_{22} = G_{21} = G_{12} = 0$.

## APPENDIX B: DIAGONALIZATION OF $\underline{\underline{\mathbf{LC}}}$ AND REAL EIGENVALUES OF LOSSLESS CTL

Consider a lossless, uniform CTL system made by two TLs, described by its per-unit length inductance and capacitance $2\times 2$ matrices, $\underline{\underline{\mathbf{L}}}$ and $\underline{\underline{\mathbf{C}}}$, respectively. They are strictly symmetric, and positive definite matrices following energy conservation requirements [42], therefore they possess real eigenvalues, however in general they are not require to commute, i.e., $\underline{\underline{\mathbf{LC}}} \neq \underline{\underline{\mathbf{CL}}}$. The four CTL eigenvalues (wavenumbers) are obtained as a solution of the polynomial $\det\left[\omega^2 \underline{\underline{\mathbf{LC}}} - k^2 \underline{\underline{\mathbf{1}}}\right] = 0$ or $\det\left[\omega^2 \underline{\underline{\mathbf{CL}}} - k^2 \underline{\underline{\mathbf{1}}}\right] = 0$, with $k$ being the wavenumber. The solution of such eigensystem for $k^2(\omega)$ can be written in terms

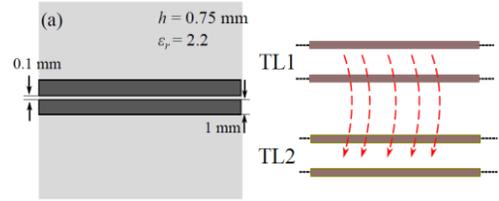

Fig. A1. Geometry of the microstrip configuration adopted in this paper (ground plane not shown here). The homogenous coupled microstrip lines develop an EPD when gain and loss are coexisting. On the right, the equivalent CTL is depicted where the red dashed arrows represent coupling and/or power transfer from the TL with gain to the one without gain.

of the inductance and capacitances matrix entries as assumed in (5) as $k_{1,2}^2(\omega) = \frac{1}{2}\omega^2\left[(L_{11}C_{11} + L_{22}C_{22} - 2L_mC_m) \pm \beta^2\right]$, where

$\beta^4 = \left[(C_{11}L_{11} - C_{22}L_{22})^2 + 4(C_mL_{11} - L_mC_{22})(C_mL_{22} - L_mC_{11})\right]$.

We shall prove that such lossless CTL never supports an EPD. Note that if $k_1^2(\omega) \neq k_2^2(\omega)$, i.e., there are two distinct eigenvalues, the associated eigenvectors must be independent (Ch. 7 in [48]) and therefore no EPDs exists in this case. However, for the case when $k_1^2(\omega) = k_2^2(\omega)$ which is necessary condition for an EPD, we will prove that such lossless system (with no loss and no gain) can only have independent eigenvectors when $\omega > 0$, i.e., no EPD exists for any frequency except at zero frequency. It is obvious that in the special case when $\underline{\underline{\mathbf{L}}}$ and $\underline{\underline{\mathbf{C}}}$ commute, the product $\underline{\underline{\mathbf{LC}}}$ in (12) is symmetric and therefore diagonalizable. Hence two independent eigenvectors are found and no EPDs occur. Therefore, let us provide a more general proof for the more involved case when $\underline{\underline{\mathbf{L}}}$ and $\underline{\underline{\mathbf{C}}}$ do not commute. For the sake of generality, let us examine whether a matrix $\underline{\underline{\mathbf{Q}}}_v$ exists to transform $\underline{\underline{\mathbf{LC}}}$ into a diagonal matrix $\underline{\underline{\mathbf{\Lambda}}} = \text{diag}(k_1^2, k_2^2)$ such that $\underline{\underline{\mathbf{\Lambda}}} = \underline{\underline{\mathbf{Q}}}_v^{-1} \underline{\underline{\mathbf{LC}}} \underline{\underline{\mathbf{Q}}}_v$. In that case, another matrix $\underline{\underline{\mathbf{Q}}}_i$ could also satisfy $\underline{\underline{\mathbf{\Lambda}}} = \underline{\underline{\mathbf{Q}}}_i^{-1} \underline{\underline{\mathbf{CL}}} \underline{\underline{\mathbf{Q}}}_i$ (recall that a sufficient condition for matrix diagonalization is the existence of a non-singular similarity transformation such as $\underline{\underline{\mathbf{Q}}}_v$ or $\underline{\underline{\mathbf{Q}}}_i$, see Ch. 7 in [48]). Now we prove that such matrices ($\underline{\underline{\mathbf{Q}}}_i$ and $\underline{\underline{\mathbf{Q}}}_v$) are non-singular, hence the matrix $\underline{\underline{\mathbf{LC}}}$ can be diagonalized. If it exists, the matrix $\underline{\underline{\mathbf{Q}}}_v$ is composed of two column eigenvectors, $\underline{\underline{\mathbf{Q}}}_v = [\mathbf{q}_{v1} | \mathbf{q}_{v2}]$ and to prove that $\underline{\underline{\mathbf{Q}}}_v$ is non-singular we just need to prove that the two regular independent eigenvectors $\mathbf{q}_{v1,2}$ exist. To do that we start writing the eigenvalue problem as $\omega^2 \underline{\underline{\mathbf{LC}}} \mathbf{q}_{v1,2} = k_{1,2}^2 \mathbf{q}_{v1,2}$ and look for eigenvalues $k_{1,2}^2$ and two eigenvectors. The problem can be rewritten as $\omega^2 \underline{\underline{\mathbf{C}}} \mathbf{q}_{v1,2} = k_{1,2}^2 \underline{\underline{\mathbf{L}}}^{-1} \mathbf{q}_{v1,2}$, and since $\underline{\underline{\mathbf{L}}}$ is symmetric and positive definite, $\underline{\underline{\mathbf{L}}}^{-1}$ is also the same. Therefore, since $\underline{\underline{\mathbf{C}}}$ is also symmetric, and positive definite, the eigenvalues $k_1^2$ and $k_2^2$ when $\omega > 0$ are always *positive and real* for such *generalized eigenvalue problem with symmetric-definite* properties (Ch. 15 and 43 in [47]). We then proceed with decomposing $\underline{\underline{\mathbf{L}}}^{-1}$ in terms of a lower triangular matrix $\underline{\underline{\mathbf{P}}}$ as $\underline{\underline{\mathbf{L}}}^{-1} = \underline{\underline{\mathbf{P}}} \underline{\underline{\mathbf{P}}}^T$ (p.34 in [61]). Accordingly, the eigenvalue problem transforms to $\underline{\underline{\mathbf{D}}} \mathbf{z}_{v1,2} = k_{1,2}^2 \mathbf{z}_{v1,2}$ in which $\underline{\underline{\mathbf{D}}} = \omega^2 \underline{\underline{\mathbf{P}}}^{-1} \underline{\underline{\mathbf{C}}} (\underline{\underline{\mathbf{P}}}^T)^{-1}$ and $\mathbf{z}_{v1,2} = \underline{\underline{\mathbf{P}}}^T \mathbf{q}_{v1,2}$. Since $\underline{\underline{\mathbf{D}}}$ is symmetric and real, two eigenvectors $\mathbf{z}_{v1,2}$ exist, and consequently $\mathbf{q}_{v1,2}$ are respectively linearly independent eigenvectors (see details in p.34 in [61]). Thus the matrix $\underline{\underline{\mathbf{ZY}}}$ is always diagonalizable (i.e., it cannot exhibit EPDs) in a lossless CTL; even when eigenvalues are repeated, i.e., when $k_1^2(\omega) = k_2^2(\omega)$. The special case when $k_1^2 = k_2^2 = 0$ only occurs when $\omega = 0$ and yields vanishing eigenvectors which is a trivial scenario.